\documentclass[12pt,preprint]{aastex}

\shorttitle{AIC of WDs}
\shortauthors{Abolimiti \& Li.}

\begin{document}

\title{Formation of Binary Millisecond Pulsars by Accretion-Induced Collapse
of White Dwarfs under Wind-Driven Evolution}
\author{Iminhaji Ablimit\altaffilmark{1,2} and Xiang-Dong Li\altaffilmark{1,2}}
\altaffiltext{1}{Department of Astronomy, Nanjing University,
Nanjing 210046, China} \altaffiltext{2}{Key Laboratory of of Modern
Astronomy and Astrophysics, Ministry of Education, Nanjing 210046,
China}

\begin{abstract}
Accretion-induced collapse of massive white dwarfs (WDs) has been
proposed to be an important channel to form binary millisecond
pulsars (MSPs). Recent investigations on thermal timescale mass
transfer in WD binaries demonstrate that the resultant MSPs are
likely to have relatively wide orbit periods ($\gtrsim 10$ days).
Here we calculate the evolution of WD binaries taking into account
the excited wind from the companion star induced by X-ray
irradiation of the accreting WD, which may drive rapid mass transfer
even when the companion star is less massive than the WD. This
scenario can naturally explain the formation of the strong-field
neutron star in the low-mass X-ray binary 4U 1822$-$37. After AIC
the mass transfer resumes when the companion star refills its Roche
lobe, and the neutron star is recycled due to mass accretion. A
large fraction of the binaries will evolve to become binary MSPs
with a He WD companion, with the orbital periods distributed between
$\gtrsim 0.1$ day and $\lesssim 30$ days, while some of them may
follow the cataclysmic variable-like evolution towards very short
orbits. If we instead assume that the newborn neutron star appears
as an MSP and part of its rotational energy is used to ablate its
companion star, the binaries may also evolve to be the redback-like
systems.

\end{abstract}


\section{Introduction}
Theoretically a neutron star (NS) can be formed in three different
ways: core-collapse supernova (CCSN) of a massive star, electron
capture supernova (ECSN) of an intermediate-mass star, and
accretion-induced collapse (AIC) of a massive white dwarf (WD)
\citep[][and references therein]{vdh09}. AIC may occur either
through rapid mass transfer onto an ONeMg WD in a binary from a
non-degenerate companion star, or through merger of two WDs in a
compact binary, and the evolutionary processes are similar to those
in the single and double degenerate scenarios for type Ia
supernovae, respectively. In the former case the ONeMg  WD may
retain the transferred H- and/or He-rich material by stable nuclear
burning and grow to the Chandrasekhar mass \citep{Iv04}. Electrons
are then captured by {\rm Mg} and {\rm Ne}, heating the
surroundings. However, the energy released by the O+Ne deflagration
is too small to cause an explosion of the tightly bound core
\citep{Mi80}. Further electron capture eventually leads to
gravitational core collapse to form an NS without an SN
\citep{Ca80,No91}.

AIC has been proposed as an alternative channel to form millisecond
pulsars (MSPs) besides the standard recycling scenario with
core-collapse NSs \citep{ch87,Mi87,ku88,ba90}. MSPs are old radio
pulsars with spin periods less than 20 ms. Most of them are found in
binary systems, and their magnetic fields ($\sim 10^{8}-10^{9}$ G)
are significantly lower than those  ($\sim 10^{11}-10^{13}$ G) of
ordinary pulsars \citep{Lo08}. In the standard recycling model, MSPs
are thought to be the descendants of low-mass X-ray binaries
(LMXBs)\footnote{Part of the LMXBs may have evolved from
intermediate-mass X-ray binaries \citep[e.g.,][]{p02}.}, in which
NSs have accreted sufficient mass and angular momentum from the
companion star via Roche lobe overflow (RLOF), and been spun up to
millisecond periods \citep[][for reviews]{Bh91,Tauris2006}. A
longstanding problem for the recycling scenario is the discrepancy
between the birth rates of Galactic LMXBs and MSPs, as first noticed
by \citet{ku88}. This problem has been tackled by many authors both
observationally and theoretically
\citep[e.g.,][]{c89,n90,c94,i95,Lo95,cc97,ly98,wg98,p03,s07,fw07,d10,Hu10}.
Most recent works on the Galactic MSP population show that the birth
rate problem is still present \citep{g13,l13}.

The advantage of the AIC scenario is that it might maintain a
sufficiently high formation rate to account for MSPs
\citep[e.g.,][]{Hu10}.
However, MSPs formed via AIC are actually difficult to distinguish
from those evolved from LMXBs, since the subsequent mass transfer
after the AIC event proceeds in a way similar as in the recycling
scenario \citep{Su00,Tau13}. The AIC scenario is especially favored
for the pulsars in globular clusters which have characteristic ages
significantly less than the ages of the clusters, suggesting that
they were formed very recently with a very small kick imparted on
the newly born NS \citep{ly96,bo11}. AIC is also invoked to explain
the strong-field pulsars with a He WD companion
\citep[e.g.,][]{Ta1986} or strong-field accreting NSs in LMXBs
\citep[e.g.,][]{van97,xu09}, which seem to have experienced
extensive mass accretion, and should have very weak fields according
to the recycling scenario.

In a recent work, \citet{Tau13} investigated the binary evolution
leading to AIC to examine if NSs formed in this way can subsequently
be recycled to form MSPs. It was found that this scenario is
possible for systems with companion stars that are either
main-sequence (MS) stars, giants stars, or He stars. The first type
of companion stars lead to fully recycled MSPs with He WD
companions, whereas the other two types of donors lead to more
mildly recycled pulsars with mainly CO WD companions. For MSP/He WD
binaries the orbital periods are predicted to lie between about 10
days and about 60 days, consistent with \citet{Hu10}. Since
observations of MSPs reveal an orbital period range between a few
hours and up to 1000 days, one needs to know whether MSPs with
shorter orbital periods can also be formed via AIC.

In this paper we explore a supplementary AIC channel to the MSP
formation taking into account the influence of irradiation-excited
wind in accreting WD binaries \citep{van1998,King1998}. Our study is
similar to \citet{Tau13}, but the evolutionary processes are
considerably different. In particular our calculations can reproduce
MSP/He WD binaries with relatively short orbital periods (less than
a few days), and the LMXBs containing a strong-field NS like 4U
1822$-$37. We also show that the evolutions can lead to the
formation of redbacks.

This paper is arranged as follows. We describe the model assumptions
in section 2. Our calculated results of pre- and post-AIC evolution
are shown section 3, which are also compared with observations of
binary MSPs. We discuss the possible implications of our results and
conclude in section 4.

\section{The wind-driven model}
\label{sec:model} The growth of the WD mass in a binary requires
that the accreted material can be stably burned on the surface of
the WD, and this usually occurs when the companion star, if it is an
MS star, is more massive than the WD, so that mass transfer can
proceed on a thermal timescale \citep{va92}. In this case the
accreting WD often appears as a supersoft X-ray sources
\citep[SSSs,][]{Kah1997}. However, the observational properties of
some SSSs including RX J0439.8$-$6809, 1E0035.4$-$7230, RX
J0537.7$-$7034, and CAL 87 \citep{Schmidtke1996,ste06,Oliveira2007},
as well as the recurrent Nova T Pyxidis \citep{kn00}, seem not to
fit in the classical picture. Most of them have orbital periods of a
few hours, smaller than expected for thermal timescale mass transfer
\citep{King2001}, and the companion stars are therefore less massive
than the WDs, but can still maintain a high mass transfer rate.

Another challenge comes from the dipping LMXB 4U 1822$-$37
\citep{mason82,cowley03}. The companion star in 4U 1822$-$37 is of
low mass ($\sim 0.44 - 0.56 M_{\sun}$) \citep{munoz05}. X-ray and
optical light curves indicate an orbital period $\sim 5.7$ hr
\citep{mason82,burderi10}. Recent {\it Suzaku} observation detected
a cyclotron resonance scattering feature at an energy of $33(\pm 2)$
keV, implying that the NS in this source has a strong magnetic field
of $2.8\times 10^{12}$ G \citep{sasano14}. Since the NS must have
accreted at least a few tenths $M_{\sun}$ matter from its companion,
which would significantly reduce the field as in typical LMXBs, an
AIC model seems to be the most likely explanation for its current
strong field. In this model the renewed mass transfer should not
have lasted long time, so that the companion star and the orbit have
not changed considerably since the AIC, but the short orbital period
and small companion mass are both inconsistent with the traditional
SSS expectation.

A possible solution to the above puzzles is the self-excited wind
model suggested by \citet{van1998} and \citet{King1998}. They argue
that perhaps in all WD binaries the soft X-ray radiation from an
accreting WD may lead to a strong stellar wind from the heated side
of the companion star. If the wind takes away the specific angular
momentum of the companion from the binary, mass transfer will be
driven at a rate comparable with the wind loss rate. The relation
between the  mass transfer rate $\dot{M}_{\rm tr}$ and the wind loss
rate $\dot{M}_{\rm w}$ obeys
\begin{equation}
\dot{M}_{\rm w}\simeq (3.5\times 10^{-7}\,M_\sun\,{\rm yr}^{-1})
(\frac{M_2}{M_\sun})^{5/6}(\frac{M}{M_\sun})^{-1/3} (\eta_{\rm
s}\eta_{\rm a})^{1/2}\phi (\frac{\dot{M}_{\rm
tr}}{10^{-7}\,M_\sun\,{\rm yr}^{-1}})^{1/2},
\end{equation}
for $M_2\lesssim M_{\rm WD}$; and
\begin{eqnarray}
\dot{M}_{\rm w}\simeq (3.5\times 10^{-7}\,M_\sun\,{\rm yr}^{-1})
(\frac{M_2}{M_\sun})^{0.95}(\frac{M}{M_\sun})^{-1/3} (\frac{M_{\rm
WD}}{M_\sun})^{-0.12}(\eta_{\rm s}\eta_{\rm a})^{1/2}\phi
(\frac{\dot{M}_{\rm tr}}{10^{-7}\,M_\sun\,{\rm yr}^{-1}})^{1/2},
\end{eqnarray}
for $M_2\gtrsim M_{\rm WD}$. Here $M_{\rm WD}$, $M_2$, and
$M=M_1+M_2$ are the WD mass, the companion mass, and the total mass,
respectively; $\eta_{\rm s}$ measures the efficiency of the WD's
spectrum in producing ionizing photons normalized to the case of
supersoft X-ray temperatures with a magnitude of $10^5$ K,
$\eta_{\rm a}$ measures the luminosity per gram of matter accreted
relative to the value for H shell burning, and $\phi$ is an
efficiency factor parameterizing the fraction of the companion's
irradiated face and the fraction of the wind mass escaping the
system.

Considering the irradiation-excited winds from the companion star
and its effect on the binary evolution, we investigate the mass
transfer processes of a binary consisting of an ONeMg WD and an MS
companion star with an updated version of Eggleton's stellar
evolution code \citep{Eggleton1971,Eggleton1973}. In addition,
angular momentum loss caused by gravitational wave radiation
\citep{ll75} and magnetic braking \citep{vz81,r83} is also included
in the calculation. We explore the parameter space of the initial
binaries for AIC, and, if NSs are formed in this way, the properties
of the resultant binaries.

During accretion the growth of the WD mass
is associated with the accumulation efficiencies of H- and He-rich matter,
\begin{equation}
\dot{M}_{\rm WD}=\eta_{\rm H}\eta_{\rm He}\dot{M}_{\rm tr},
\end{equation}
where $\eta_{\rm H}$ and $\eta_{\rm He}$ represent the fraction of
the transferred H- and He-rich matter from the companion that
eventually burns into He- and C-rich matter and stays on the WD,
respectively. Here we fit the numerical results of
\citet{Pralnik1995} and \citet{Yaron2005} for the H mass
accumulation efficiency $\eta_{\rm H}$, and adopt the prescriptions
in \citet{Ka2004} for the He mass accumulation efficiency $\eta_{\rm
He}$. If the WD mass reaches the Chandrasekhar limit ($M_{\rm Ch} =
1.38M_\sun$), we assume that the WD collapses to be a NS with a
gravitational mass of $1.25M_\sun$ (termed as the Chandrasekhar
model). During this process $0.13M_\sun$ mass is assumed to convert
into the binding energy. The sudden mass loss makes the orbit wider
and the temporary detachment of the RL. The relation between the
orbital separations ($a_0$ and $a$) just before and after the
collapse is \citep{Ve90}
\begin{equation}
\frac{a}{a_0} = \frac{M_{\rm WD} + M_2}{M_{\rm NS} + M_2},
\end{equation}
where $M_{\rm NS}$ is the NS mass. We also assume that there is no kick
received by the newborn NS. Investigations by \citet{Hu10} and \citet{Tau13}
have shown that including a kick velocity with a dispersion of 50 kms$^{-1}$
does not seriously affect the final results.

The observed super-luminous SNe Ia hint that there may be
super-Chandrasekhar mass WD progenitors \citep{ho06, hi07, sc10}.
\citet{yo04, yo05} found that rapid rotation allows a massive WD to
continue accreting when the accretion rate $>3\times 10^{-7}
{M_\sun}\,{\rm yr}^{-1}$, and there would not be the central C
ignition even if its mass exceeds $M_{\rm Ch}$. We accordingly
assume that a super-Chandrasekhar mass WD can exist and it undergoes
a collapses when $\dot{M}_{\rm tr}<3\times 10^{-7} {M_\sun}\,{\rm
yr}^{-1}$ so that there is no differential rotation to support the
WD (termed as the super-Chandrasekhar model).

After AIC the companion star will refill its RL due to angular
momentum loss or nuclear/thermal evolution. Mass transfer is then
resumed, and the post-AIC binary evolves as an LMXB. The interplay
between angular momentum loss and nuclear expansion of the donor
leads to the so-called bifurcation periods ($P_{\rm bif}$). LMXBs
with orbital periods shorter or longer than $P_{\rm bif}$ will form
converging or diverging systems, respectively \citep{py88, py89}.

Following \citet{Tau13} we assume that the mass transfer is
nonconservative and Eddington limited. The NS accretion rate is
described by the following equation,
\begin{equation}
\dot{M}_{\rm NS} = (|\dot{M}_2| - {\rm max}[|\dot{M}_2|-\dot{M}_{\rm Edd}, 0])\cdot e_{\rm a}\cdot k_{\rm d},
\end{equation}
where $\dot{M}_{\rm Edd}$ ($\simeq 2\times10^{-8}M_\sun\,{\rm
yr}^{-1}$) is the Eddington accretion rate, $e_{\rm a}$ is the
fraction of the transferred matter that remains on the NS, and
$k_{\rm d}$ is a factor denoting the ratio of gravitational mass to
rest mass of the accreted matter. In this paper we take $e_{\rm
a}\cdot k_{\rm d} = 0.35$.
%

\section{Results}
We have performed numerical calculation of the evolution of a grid
of binaries consisting of an ONeMg WD of initial mass $M_{\rm
WD,i}=1.2 M_\odot$ and  an MS companion star of initial mass $M_{\rm
2,i}=0.6-7\, M_\odot$ with both Pop I and II compositions under the
irradiation-excited wind-driven model. The detailed calculated
results are described as follows.

Figure 1 shows the distribution of the initial companion's mass
$M_{\rm 2,i}$ versus the initial orbital period $P_{\rm orb,i}$, in
which the binaries can evolve to AIC. The left, middle, and right
panels correspond to the Chandrasekhar model with metallicities
$Z=0.02$ and 0.001, and the super-Chandrasekhar model with $Z=0.02$,
respectively. It is seen from the left and middle panels that, while
the upper limit of the donor mass is about $3-3.5M_\sun$, similar to
that in the standard SSS model, the lower limit can go down to $\sim
0.7-0.8\,M_\odot$. The initial orbital periods range from about
${0.2}\,{\rm day}$ to about $4\,{\rm days}$. Compared with the
results without the wind considered \citep{Li97,Tau13} in which the
lower limit of the donor mass $\sim 2\,M_{\sun}$, the donor mass
extends to smaller mass ($\lesssim 1\,M_{\sun}$), and the orbital
period distribution becomes narrower. The reason is that here the
mass transfer is mainly driven  by the excited wind when $M_2$ is
less than $M_{\rm WD}$, rather than the thermal evolution of the
companion star. Meanwhile, the wind can somewhat stabilize the mass
transfer when $M_2$ is larger than $M_{\rm WD}$. Outside the
confined regions either the mass transfer becomes dynamically
unstable, or the WD can never reach the Chandrasekhar limit because
of relatively low mass transfer rate. In the right panel, the WD is
assumed to rotate rapidly and continue accreting matter when its
mass exceeds the Chandrasekhar limit. When the mass transfer rate
becomes below $3\times10^{-7} {M_\sun}{\rm yr}^{-1}$, it slows down
and begins to collapse. In this case, systems that can successfully
evolve to AIC have companions of initial masses $\sim 1 -
3.5M_\sun$, and initial orbital periods $\sim 0.25 -3.5$ days. The
parameter space is smaller than in the former two cases because
generally more massive donors are required to sustain sufficient
matter to form a super-Chandrasekhar WD. In the figure the red lines
divide the final evolutionary products of the binaries. Systems
above the red line will evolve to become NS/WD binaries. The
companion stars in the regions below the red line are of low mass
and in short orbits after AIC, and they stay in the MS phase within
the Hubble time.

In Figure 2, we show the distribution of the companion mass $M_{\rm
2}$ versus the orbital period $P_{\rm orb}$ at the moment of AIC.
The left, middle, and right panels correspond to those in Fig.~1,
respectively. The most striking feature is that the orbital periods
occupy in a small range $\sim 0.3 - 1.7$ days, and the companion
mass goes down to $\sim 0.2-0.3M_\sun$ for the Chandrasekhar model.
The low mass of the companion star and the short orbital period
mainly originate from the mass and angular momentum loss related to
the irradiation-excited wind. The WD accretes more mass from the
donor in the super-Chandrasekhar model. Therefore, the range of the
companion mass ($\sim 0.4-1.1M_\sun$) when AIC occurs is smaller,
but the orbital period range ($\sim 0.3 - 2.9$ days) is a bit
larger.

Figures 3 and 4 display four examples of the evolution of the mass
transfer rate, the orbital period, the companion mass, and the WD/NS
mass for binaries with different initial parameters in the pre- and
post-AIC phases, respectively. Here we adopt the Chandrasekhar model
with $Z=0.02$.

In the top panel of Fig.~3 we take $M_{\rm 2,i}=1.7M_\sun$ and
$P_{\rm orb,i}=1.25$ days. In such a binary the traditional
thermal-timescale mass transfer is not rapid enough for stable H
burning. However, its rate is enhanced by the irradiation-excited
wind up to a few $10^{-7} {M_\sun}\,{\rm yr}^{-1}$, thus the WD can
grow in mass to $M_{\rm Ch}$. The orbital period decreases along
with the mass transfer. In the second panel we take $M_{\rm
2,i}=2.54M_\sun$ and $P_{\rm orb, i}=1.95$ days. Since the donor
mass is considerably larger than the WD mass, although the
irradiation-excited wind also works, the mass transfer actually
proceeds on a thermal timescale, similar as in typical SSSs. In the
third panel we take $M_{\rm 2,i}=1.8M_\sun$ and $P_{\rm orb, i}=1.0$
day. The binary evolution seems to be similar to that in the top
panel before AIC, but differs after AIC as shown in Fig.~4. In the
bottom panel we take $M_{\rm 2,i}=1.5M_\sun$ and $P_{\rm orb,
i}=0.75$ day, which are below the red line in Fig.~1. This binary
will evolve towards to an very compact system after AIC.

Figure 4 shows the post-AIC mass transfer processes in the four
binaries discussed above. The binaries evolve in a similar way as
low- and intermediate-mass X-ray binaries. In the first one (in the
top panel), the mass transfer initiates when the orbital period
$\sim 0.34$ day and the companion mass $\sim0.48M_\sun$. These
values are roughly in accord with the donor mass and the orbital
period of the LMXB 4U 1822$-$37. Hence if the newborn NS possesses a
strong magnetic field ($\sim 10^{12}$ G), the evolutionary sequence
in the top panels of Figs.~3 and 4 provides a possible formation
path of 4U 1822$-$37. The orbital period increases with mass
transfer to $\sim 3.07$ days when the donor loses its envelope and
leaves a He WD. About $0.197M_\sun$ mass is accreted by the NS,
which should be recycled to be an MSP. In the second case, the mass
transfer begins at a longer orbital period ($\sim 1.2$ days). The
orbital period increases to $\sim 30$ days when the donor evolves to
become a WD. The NS accretes a large amount of mass (${\Delta M}\sim
{0.54M_\sun}$), and a significant decay of its magnetic field is
also expected. In the third panel the mass transfer initiates when
the orbital period $\sim 0.30$ day. That is very close to $P_{\rm
bif}$ so the final orbital period does not change much. The
companion star finally becomes a $\sim 0.16 M_{\sun}$ He WD, and the
NS is recycled by accreting about  $0.18M_\sun$ mass. The last
binary evolves all the way to be a very compact X-ray binary. The NS
continues accreting mass when the donor star keeps on the MS, and
the orbital period decreases down to $\sim 0.066$ day when the donor
mass is lower than $0.1 M_\sun$. The fate of the binary may be a
black widow system.

In Table 1 we list the calculated parameters of selected binary
evolutionary sequences. In most cases the NSs can accrete
$>0.05\,M_{\sun}$ mass during the post-AIC mass transfer phase,
which seems to be sufficient to reduce the NS fields to $\lesssim
10^9$ G and accelerate the NS spin periods to $\lesssim 20$ ms.
However, this is strongly dependent on the (unknown) mass transfer
efficiency, and there is mounting evidence that NSs may accrete a
small fraction of the transferred mass in the evolution of LMXBs
\citep[e.g.,][]{j05,a12}.

Figure 5 shows the the distribution of the produced NS/WD binaries
in the orbital period ($P_{\rm orb,f}$) vs. the WD mass ($M_{\rm
2,f}$) diagram. The solid, dotted and dashed lines represent the
results of the Chandrasekhar model with $Z=0.02$ and 0.001, and of
the super-Chandrasekhar model with $Z=0.02$, respectively. The
observed binary pulsars with a WD companion are also plotted in
dots, with different colors denoting the range of the pulsar
magnetic fields (data are taken from the ATNF pulsar
catalogue\footnote{http:www.atnf.csiro.au/research/pulsar/psrcat}):
red, blue, and green colors are for $B<10^{10}$ G, $10^{10}$ G
$<B<10^{12}$ G, and $B>10^{12}$ G, respectively. They roughly
correspond to recycled, mildly recycled, and non-recycled pulsars,
respectively. It is seen that the predicted orbital periods of NS/WD
binaries are distributed between $\gtrsim 0.1$ day and $\lesssim 30$
days, and the WD masses are between $\sim 0.15M_\sun$ and $\sim
0.45M_\sun$, compatible with a large fraction of the known MSP/He WD
binaries\footnote{Note that the WD masses for binary pulsars are
plotted with the median masses by assuming an orbital inclination
angle of $60\degr$ and a pulsar mass of $1.35\,M_{\sun}$. We do not
plot the error bars for the clarity of the figure.}. However, it is
still unable to account for some peculiar systems like PSR
1831$-$00, which has a strong ($7.5\times10^{10}$ G) magnetic field,
a short (1.81 days) orbit, and a very low-mass ($0.075M_\sun$) WD
companion \citep{Su00}.

\section{Discussion and conclusions}

With population synthesis calculations, \cite{Hu10} suggested that,
while both the CCSN and AIC channels lead to populations of X-ray
binaries and binary MSPs at the end of the accretion phase, the
birthrates of binary MSPs via AIC  are comparable to or even exceed
those for CCSNe, and it appears to be the major channel for the
pulsars in long-period ($>$ a few days) systems with He WD
companions under certain model assumptions. These conclusions are
further confirmed by \citet{Tau13} with detailed evolutionary
calculations. They showed that MSPs formed via AIC and which have He
WD companions generally have $P_{\rm orb}$ between 10 and 60 days.
However, as pointed out by \cite{Hu10},  both the AIC and CCSN
channels have a problem producing the observed binary MSPs with 0.1
day $\leq P_{\rm orb} \leq 5$ days.

In this paper we investigate the formation of MSP/He WD binaries
with an AIC origin, taking into account the effect of the
irradiation-excited wind. Because of the angular momentum loss
associated with the wind mass loss, the initial orbital periods of
successful systems are always within a few days. The resultant
MSP/He WD binaries are inclined to have orbital periods between
$\gtrsim 0.1$ day and $\lesssim 30$ days, and He WDs with masses
between $\sim 0.15M_\sun$ and $\sim 0.45M_\sun$. If we combine the
results of \cite{Hu10} and \citet{Tau13} with this work, it seems
that the AIC channel can at least cover the majority of the orbital
period range of the observed MSP/WD binaries.

The predicted orbital period and WD mass distributions of the binary
pulsar systems depicted in Fig.~5 are also in broad agreement with
those in previous investigations on L/IMXBs
\citep{r95,t99,p02,l11,s14b,j14,i14}. For binaries with low-mass He
WDs, lower metallicities tend to result in shorter orbital periods,
similar as in \citet{j14}. However, since MSPs with He WD companions
in very compact binaries can also be accounted for by LMXB evolution
if the progenitor binary experienced very late Case A mass transfer
\citep{s14b,j14,i14}, it is difficult to distinguish the formation
channels for MSPs only from their currently measured parameters.
Besides (partially) alleviating the birth rate discrepancy between
MSPs and LMXBs, the AIC channel under wind-driven evolution may be
preferred for the formation of the strong-field NSs in ``old"
binaries with short orbital periods. The NS in 4U 1822$-$37 as
mentioned before is one example. The binary radio pulsar PSR
B1718$-$19 in the globular cluster NGC 6342 \citep{l93} could be
another example. It is a young, long-spin period ($\sim 1$ s)
pulsar, with a characteristic age of 10 Myr and a magnetic field of
$1.5\times 10^{12}$ G. The origin of such apparent young objects in
very old systems has not been understood. If PSR B1718$-$19 was
formed through AIC recently \citep[e.g.,][]{ly96}, its 6.2 hr
orbital period can be accounted for by the irradiation-excited wind
during the previous mass transfer phase.

The magnetic field of a newborn NS after AIC depends on the property
and the accretion history of the WD, as well as the complicated
field decay/generation mechanisms \citep[][and references
therein]{fw07}, so that it is difficult to precisely predict the
field distribution. If the WD originally has a very weak field, the
NS may be born with rapid rotation (milliseconds in periods) and a
low field ($\sim 10^8$ G), if we assume the magnetic flux is
conserved. During the AIC some baryonic mass is abruptly lost to its
binding energy so that the orbit expands and the companion star is
detached from its RL. The mass transfer terminates and the NS
appears as an MSP, which may be able to ablate/evaporate the
companion with its high-energy radiation and particles, leading to
the redback-like systems \citep[see][for a review on redbacks]{r13}.
This turning-on of an MSP activity was previously assumed to occur
during the LMXB evolution when the mass transfer rate is temporarily
decreased \citep{c13,b14}. The AIC scenario for the redback
formation was recently proposed by \citet{s14a}. They showed that
the subsequent evolution is determined by orbital angular momentum
loss owing to gravitational radiation and magnetic braking and
ablation of the companion star at a rate of \citep{s92}
\begin{equation}
-\dot{M}_2=\frac{fL_{\rm psr}}{2v_{\rm 2,
esc}^2}\left(\frac{R_2}{a}\right)^2,
\end{equation}
where $L_{\rm psr}$ is the MSP's spin-down luminosity (taken a
typical value of $1.5\times 10^{34}$ ergs$^{-1}$), $f$ an efficiency
parameter denoting the fraction of the PSR's luminosity that is used
to ablate the companion, $v_{\rm 2, esc}$ the escape velocity of a
thermal wind from the surface of the companion, $R_2$ the
companion's radius, and $a$ the binary separation. It was shown that
if $f > 0.12$, ablation is strong enough to overcome the pull of
magnetic braking immediately after AIC and the systems evolve to
longer orbital periods without the occurrence of a second RLOF. In
\citet{s14a}, all the initial systems have a $1.2 M_{\sun}$ WD and a
$1.0 M_{\sun}$ donor with a 0.3 day orbital period. However, it has
been already known that the evolution of such a binary cannot lead
to the formation of a Chandrasekhar-mass WD, because the mass
transfer driven by magnetic braking is too low to allow stable H and
He burning \citep[e.g.,][]{Li97,Iv04,Tau13}. Aided with an
irradiation-excited wind it is able to evolve to AIC as shown in
this work. We have calculated the binary evolution after AIC
assuming that the NSs are born as MSPs. In Fig.~6 we illustrate the
evolutionary tracks for six example binary pulsars. Here the green,
black, and red lines are for the cases with $f=0.1$, 0.45, and 0.8,
respectively. The blue triangles show the positions of known
redbacks. Figure 6 confirms that it is possible to account for
redbacks at both small and large companion masses by changing the
value of $f$ (or $L_{\rm psr}$) within a proper range.

Finally it should be noted that the conditions for the wind-driven
evolution are not well understood, so the birthrate of MSPs via
wind-driven AIC currently cannot be confidently estimated. Although
there is evidence that there is or has been rapid mass transfer in
short-period WD binaries with a low-mass companion star, most of the
known such WD binaries are ordinary cataclysmic variables,
suggesting that the wind-driven case might not be popular, and its
occurrence requires some special conditions \citep[see][for a
discussion]{King1998}. Obviously a thorough investigation on this
subject will be of great value not only for AIC and SNe Ia, but also
for the overall evolution of cataclysmic variables.

\acknowledgments This work was supported by the Natural Science
Foundation of China under grant number 11133001 and 11333004, and
the Strategic Priority Research Program of CAS under grant No.
XDB09000000.


\begin{thebibliography}{99}

\bibitem[\protect\citeauthoryear{Antoniadis et al.}{2012}]{a12}
Antoniadis, J., van Kerkwijk, M. H., Koester, D., et al. 2012,
\mnras, 423, 3316
\bibitem[\protect\citeauthoryear{Bailyn \& Grindlay}{1990}]{ba90}
Bailyn, C. D., \& Grindlay, J. E. 1990, \apj, 353, 159
\bibitem[\protect\citeauthoryear{Benvenuto et al.}{2014}]{b14}
Benvenuto, O. G., De Vito, M. A., \& Horvath, J.E., 2014, \apj, 786,
L7
\bibitem[\protect\citeauthoryear{Bhattacharya \& van den Heuvel}{1991}]{Bh91}
Bhattacharya, D., \& van den Heuvel, E. P. J.  1991, Phys. Rep, 203, 1
\bibitem[\protect\citeauthoryear{Boyles et al.}{2011}]{bo11}
Boyles, J., Lorimer, D. R., Turk, P. J., et al. 2011, \apj, 742, 51
\bibitem[\protect\citeauthoryear{Burderi et al.}{2010}]{burderi10}
Burderi, L., Di Salvo, T., Riggio, A., et al. 2010, \aap, 515, A44
\bibitem[\protect\citeauthoryear{Camilo et al.}{1994}]{c94}
Camilo, F., Thorsett, S. E., \& Kulkarni, S. R. 1994, \apjl, 421,
L15
\bibitem[\protect\citeauthoryear{Cordes \& Chernoff}{1997}]{cc97}
Cordes, J. M., \& Chernoff, D. F., 1997, \apj, 482, 971
\bibitem[\protect\citeauthoryear{Cot\'e \& Pylyser}{1989}]{c89}
Cot\'e, J. \& Pylyser, E.H.P.  1989, \aap, 218, 13
\bibitem[\protect\citeauthoryear{Canal et al.}{1980}]{Ca80}
Canal, R., Isern, J., \& Labay, J.  1980, \apjl, 241, L33
\bibitem[\protect\citeauthoryear{Chanmugam \& Brecher}{1987}]{ch87}
Chanmugam, G., \& Brecher, K. 1987, \nat, 329, 696
\bibitem[\protect\citeauthoryear{Chen et al.}{2013}]{c13}
Chen, H.-L., Chen, X., Tauris, T. M., \& Han, Z., 2013, \apj, 775,
27
\bibitem[\protect\citeauthoryear{Cowley et al.}{2003}]{cowley03}
Cowley, A. P., Schmidtke, P. C., Hutchings, J. B., \& Crampton, D. 2003, \aj, 125, 2163
\bibitem[\protect\citeauthoryear{Dai \& Li}{2010}]{d10}
Dai, H., \& Li, X.-D. 2010, Sci China Phys Mech Astron, Vol.53
Suppl.1:125, 129
\bibitem[\protect\citeauthoryear{Eggleton}{1971}]{Eggleton1971}
Eggleton, P. P. 1971, \mnras, 151, 351
\bibitem[\protect\citeauthoryear{Eggleton}{1973}]{Eggleton1973}
Eggleton, P. P. 1973, \mnras, 163, 279
\bibitem[\protect\citeauthoryear{Ferrario \& Wickramasinghe}{2007}]{fw07}
Ferrario, L., \& Wickramasinghe, D. 2007, \mnras, 375, 1009
\bibitem[\protect\citeauthoryear{Gr\'egoire \&  Kn\"odlseder}{2013}]{g13}
Gr\'egoire, T. \& Kn\"odlseder, J.  2013, \aap, 554, A62
\bibitem[\protect\citeauthoryear{Hicken et al.}{2007}]{hi07}
Hicken, M., Garnavich, P. M., Prieto, J. L., et al. 2007, \apj, 669,
L17
\bibitem[\protect\citeauthoryear{Howell et al.}{2006}]{ho06}
Howell, D. A., Sullivan, M., Nugent, P. E., et al. 2006, \nat, 443, 308
\bibitem[\protect\citeauthoryear{Hurley et al.}{2010}]{Hu10}
Hurley, J. R., Tout, C. A., Wickramasinghe, D. T. Ferrario, L.,
\& Kiel, P. D. 2010, \mnras, 402, 1437
\bibitem[\protect\citeauthoryear{Iben, Tutukov \& Yungelson}{1995}]{i95}
Iben, Jr. I., Tutukov, A.V. \& Yungelson, L. 1989, \apjs, 100, 233
\bibitem[\protect\citeauthoryear{Istrate et al.}{2014}]{i14}
Istrate, A., Tauris, T., \& Langer, N. 2014, \aap, 571, A45
\bibitem[\protect\citeauthoryear{Ivanova \& Taam}{2004}]{Iv04}
Ivanova, N., \& Taam, R. E.  2004, \apj, 105, 145
\bibitem[\protect\citeauthoryear{Jacoby et al.}{2005}]{j05}
Jacoby, B. A., Hotan, A., Bailes, M., Ord, S., \& Kulkarni, S. R.
2005, \apjl, 629, L113
\bibitem[\protect\citeauthoryear{Jia \& Li}{2014}]{j14}
Jia, K., \& Li, X.-D. 2014, \apj, 791, 127
\bibitem[\protect\citeauthoryear{Kahabka \& van den Heuvel}{1997}]{Kah1997}
Kahabka, P., \& van den Heuvel. E. P. J. 1997, \araa, 35, 69
\bibitem[\protect\citeauthoryear{Kato \& Hachisu}{2004}]{Ka2004}
Kato, M., \& Hachisu, I. 2004, \apjl, 613, L129
\bibitem[\protect\citeauthoryear{Knigge et al.}{2000}]{kn00}
Knigge, Ch., King, A. R., \& Patterson, J. 2000, \aap, 364, L75
\bibitem[\protect\citeauthoryear{King et al.}{2001}]{King2001}
King, A. R., Schenker, K., Kolb, U., Davies, M. B. 2001, \mnras, 321, 327
\bibitem[\protect\citeauthoryear{King \& van Teeseling}{1998}]{King1998}
King, A. R. \& van Teeseling, A. 1998, \aap, 338, 965
\bibitem[\protect\citeauthoryear{kitaura et al.}{2006}]{ki06}
Kitaura, F. S., Janka, H. T., \& Hillebrandt, W. 2006, \aap, 450, 345
\bibitem[\protect\citeauthoryear{Kulkarni \& Narayan}{1988}]{ku88}
Kulkarni, S. R., \& Narayan, R. 1988, \apj, 335, 755
\bibitem[\protect\citeauthoryear{Landau \& Lifshitz}{1975}]{ll75}
Landau, L. D., \& Lifshitz, E. M. 1975, in Course of Theoretical
Physics, Pergamon International Library of Science, Technology,
Engineering and Social Studies (4th rev. Engl. ed.; Oxford:
Pergamon)
\bibitem[\protect\citeauthoryear{Levin et al.}{2013}]{l13}
Levin, L.,
Bailes, M., Barsdell, R.B., Bates, D.S., et al.  2013, \mnras, 434, 1387
\bibitem[\protect\citeauthoryear{Li \& van den Heuvel}{1997}]{Li97}
Li, X.-D., \& van den Heuvel, E. P. J. 1997, \aap, 322, L9
\bibitem[\protect\citeauthoryear{Lin et al.}{2011}]{l11}
Lin, J., Rappaport, S., Podsiadlowski, P., et al. 2011, \apj, 732,
70
\bibitem[\protect\citeauthoryear{Lorimer}{1995}]{Lo95}
Lorimer, D. R. 1995, \mnras, 274, 300
\bibitem[\protect\citeauthoryear{Lorimer}{2008}]{Lo08}
Lorimer, D. R. 2008, Living Rev. Relativ., 11, 8
\bibitem[\protect\citeauthoryear{Lyne et al.}{1993}]{l93}
Lyne, A. G., Biggs, J. D., Harrison, P. A., \& Bailes, M. 1993,
Nature, 361, 47
\bibitem[\protect\citeauthoryear{Lyne et al.}{1996}]{ly96}
Lyne, A. G. Manchester, R. N., \& D'Amico, N. 1996, \apj, 460, L41
\bibitem[\protect\citeauthoryear{Lyne et al.}{1998}]{ly98}
Lyne, A. G., Manchester, R. N., Lorimer, D. R., et al. 1998, \mnras,
295, 743
\bibitem[\protect\citeauthoryear{Mason et al.}{1982}]{mason82}
Mason, K. O., Murdin, P. G., Tuohy, I. R., Seitzer, P., \& Branduardi-Raymont, G. 1982, \mnras,
200, 793
\bibitem[\protect\citeauthoryear{Michel}{1987}]{Mi87}
Michel, F. C.   1987, \nat, 329, 310
\bibitem[\protect\citeauthoryear{Miyaji et al.}{1980}]{Mi80}
Miyaji, S., Nomoto, K., Yokoi, K., \& Sugimoto, D. 1980, \pasj, 32, 303
\bibitem[\protect\citeauthoryear{Mu\~noz-Darias et al.}{2005}]{munoz05}
Mu\~noz-Darias, T., Casares, J., \& Mart\'inez-Pais, I. G. 2005, \apj, 635, 502
\bibitem[\protect\citeauthoryear{Narayan}{1990}]{n90}
Narayan, R., \& Ostriker, J. P. 1990, \apj, 352, 222
\bibitem[\protect\citeauthoryear{Nomoto \& Kondo}{1991}]{No91}
Nomoto, K., \& Kondo, K. 1991, \apj, 367, L1
\bibitem[\protect\citeauthoryear{Oliveira \& Steiner}{2012}]{Oliveira2007}
Oliveira, A. S. \& Steiner, J. E. 2012, \aap, 472, L21
\bibitem[\protect\citeauthoryear{Pfahl et al.}{2003}]{p03}
Pfahl, E., Rappaport, S., \& Podsiadlowski, P. 2003, \apj, 597, 1036
\bibitem[\protect\citeauthoryear{Podsiadlowski, Rappaport, \& Pfahl}{2002}]{p02}
Podsiadlowski, P., Rappaport, S., \& Pfahl, D.E.  2002, \apj, 565,
1107
\bibitem[\protect\citeauthoryear{Prialnik \& Kovetz}{1995}]{Pralnik1995}
Prialnik, D., \& Kovetz, A. 1995, \apj, 445, 789
\bibitem[\protect\citeauthoryear{Pylyser \& Savonije}{1988}]{py88}
Pylyser, E., \& Savonije, G. J. 1988, \aap, 191, 57
\bibitem[\protect\citeauthoryear{Pylyser \& Savonije}{1989}]{py89}
Pylyser, E., \& Savonije, G. J. 1989, \aap, 208, 52
\bibitem[\protect\citeauthoryear{Rappaport et al.}{1995}]{r95}
Rappaport, S., Podsiadlowski, P., Joss, P. C., Di Stefano, R., \&
Han, Z. 1995, \mnras, 273, 731
\bibitem[\protect\citeauthoryear{Rappaport et al.}{1983}]{r83}
Rappaport, S., Verbunt, F., \& Joss, P. C. 1983, \apj, 275, 713
\bibitem[\protect\citeauthoryear{Roberts}{2013}]{r13}
Roberts, M. S. E. 2013, in IAU Symp., 291, Neutron Stars and
Pulsars: Challenges and Opportunities After 80 Years, ed. J. van
Leeuwen (Cambridge: Cambridge Univ. Press), 127
\bibitem[\protect\citeauthoryear{Sasano et al.}{2014}]{sasano14}
Sasano, M., Makishima, K., Sakurai, S., Zhang, Z., \& Enoto, T.
2014, \pasj, 66, 35
\bibitem[\protect\citeauthoryear{Scalzo et al.}{2010}]{sc10}
Scalzo, R. A., Aldering, G., Antilogus, P., et al. 2010, \apj, 713, 1073
\bibitem[\protect\citeauthoryear{Schmidtke et al.}{1996}]{Schmidtke1996}
Schmidtke, P. C., Cowley, A. P., McGrath, T. K.,
Hutchings, J. B., Crampton, D. 1996, \aj, 111, 788
\bibitem[\protect\citeauthoryear{Smedley et al.}{2014a}]{s14a}
Smedley, L.S., Tout, A. C., \& Ferrario, L. 2014a,
arXiv:1410.8352
\bibitem[\protect\citeauthoryear{Smedley et al.}{2014b}]{s14b}
Smedley, L.S., Tout, A. C., Ferrario, L., \& Wickramasinghe, T. D.
2014b, \mnras, 437, 2217
\bibitem[\protect\citeauthoryear{Stevens et al.}{1992}]{s92}
Stevens, I. R., Rees, M. J., \& Podsiadlowski, P. 1992, \mnras, 254,
19
\bibitem[\protect\citeauthoryear{Steiner et al.}{2006}]{ste06}
Steiner, J. E., Oliveira, A. S., Cieslinski, D., \& Ricci, T. V.
2006, \aap, 447, L1
\bibitem[\protect\citeauthoryear{Story et al.}{2007}]{s07}
Story, S. A., Gonthier, P. L., \& Harding, A. K. 2007, \apj, 671,
713
\bibitem[\protect\citeauthoryear{Sutantyo \& Li}{2000}]{Su00}
Sutantyo, W., \& Li, X.-D. 2000, \aap, 360, 633
\bibitem[\protect\citeauthoryear{Taam \& van den Heuvel}{1986}]{Ta1986}
Taam, R. E., \& van den Heuvel, E. P. J. 1986, \apj, 305, 235
\bibitem[\protect\citeauthoryear{Tauris et al.}{2013}]{Tau13}
Tauris, T. M., Sanyal, D., Yoon, S. -C., \& Langer, N. 2013, \aap,
558, A39
\bibitem[\protect\citeauthoryear{Tauris \& Savonije}{1999}]{t99}
Tauris, T. M. \& Savonije, G. J. 1999, \aap, 350, 928
\bibitem[{{Tauris \& van den Heuvel} (2006)}]{Tauris2006}
Tauris T. M. \& van den Heuvel E. P. J. 2006, in Compact
stellar X-ray sources. ed. W. Lewin \& M. van der Klis
(Cambridge: Cambridge Univ. Press), 623
\bibitem[\protect\citeauthoryear{van den Heuvel}{2009}]{vdh09}
van den Heuvel, E. P. J. 2009, in Physics of Relativistic Objects in Compact Binaries:
From Birth to Coalescence, Astrophysics and Space Science Library, Vol. 359
(Springer Netherlands), p. 125
\bibitem[\protect\citeauthoryear{van den Heuvel et al.}{1992}]{va92}
van den Heuvel, E. P. J., Bhattacharya, D., Nomoto, K., \& Rappaport, S. A. 1992, \aap, 262, 97
\bibitem[\protect\citeauthoryear{van Paradijs et al.}{1997}]{van97}
van Paradijs, J., van den Heuvel, E. P. J., Kouveliotou, C., et al. 1997, \aap, 317, L9
\bibitem[\protect\citeauthoryear{van Teeseling \& King}{1998}]{van1998}
van Teeseling, A. \& King, A. R. 1998, \aap, 338, 957
\bibitem[\protect\citeauthoryear{Verbunt \& Zwaan}{1981}]{vz81}
Verbunt, F., \& Zwaan, C. 1981, \aap, 100, L7
\bibitem[\protect\citeauthoryear{Verbunt et al.}{1990}]{Ve90}
Verbunt, F. Wijers, R. A. M. J., \& Burn, H. M. G. 1990, \aap, 234, 195
\bibitem[\protect\citeauthoryear{White \& Ghosh}{1998}]{wg98}
White, N. E., \& Ghosh, P. 1998, \apjl, 504, L31
\bibitem[\protect\citeauthoryear{Xu \& Li}{2009}]{xu09}
Xu, X.-J., \& Li, X.-D. 2009, \aap, 495, 243
\bibitem[\protect\citeauthoryear{Yaron et al.}{2005}]{Yaron2005}
Yaron, O., Prialnik, D., Shara, M. M., \& Kovetz, A. 2005, \apj, 623, 398
\bibitem[\protect\citeauthoryear{Yoon \& Langer}{2004}]{yo04}
Yoon, S. -C., \& Langer, N. 2004, \aap, 419, 623
\bibitem[\protect\citeauthoryear{Yoon \& Langer}{2005}]{yo05}
Yoon, S. -C., \& Langer, N. 2005, \aap, 435, 967



\end{thebibliography}

\begin{figure}
\centering
\includegraphics[scale=0.43]{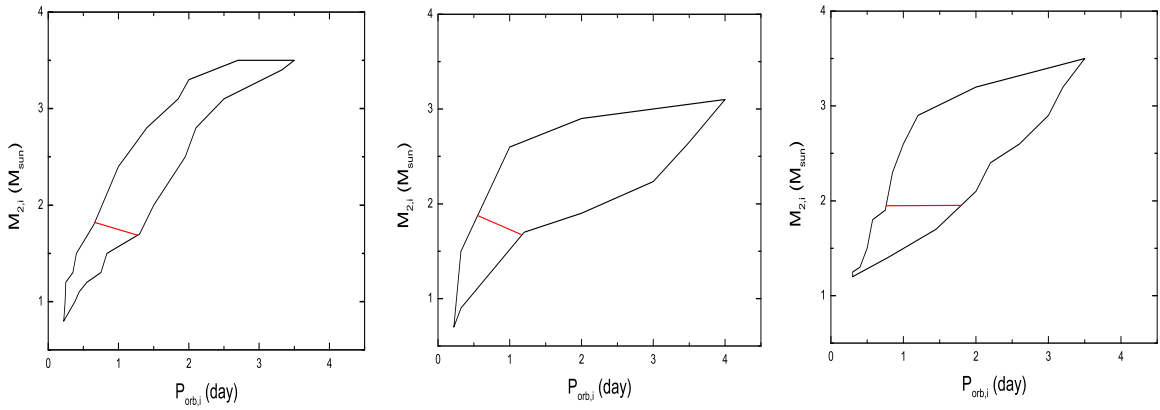}
\caption{The distributions of the initial orbital periods and the
companion masses of the binaries that can successfully evolve to AIC
in the Chandrasekhar model with metallicities $Z=0.02$ (left) and
$Z=0.001$ (middle). The right panel is for the case of
super-Chandrasekhar mass WD with $Z=0.02$. Systems above the red
line will finally evolve to be MSP/WD binaries}\label{fig:1}
\end{figure}

\clearpage

\begin{figure}
\centering
\includegraphics[scale=0.43]{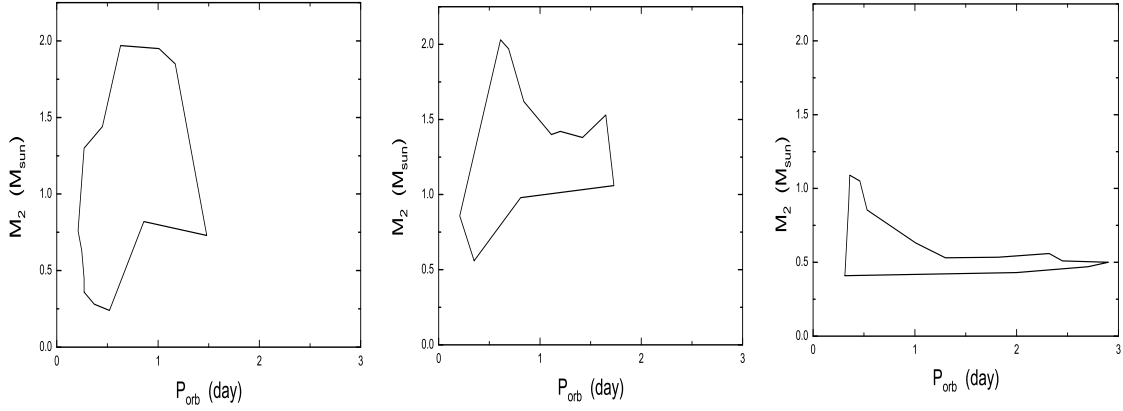}
\caption{The distributions of the orbital periods and the companion
masses of the binaries at the moment of AIC. The left, middle, and
right panels are for the Chandrasekhar model with $Z=0.02$ and
$Z=0.001$, and the super-Chandrasekhar model with $Z=0.02$,
respectively.}\label{fig:1}
\end{figure}

\clearpage

\begin{figure}
\centering
\includegraphics[scale=0.43]{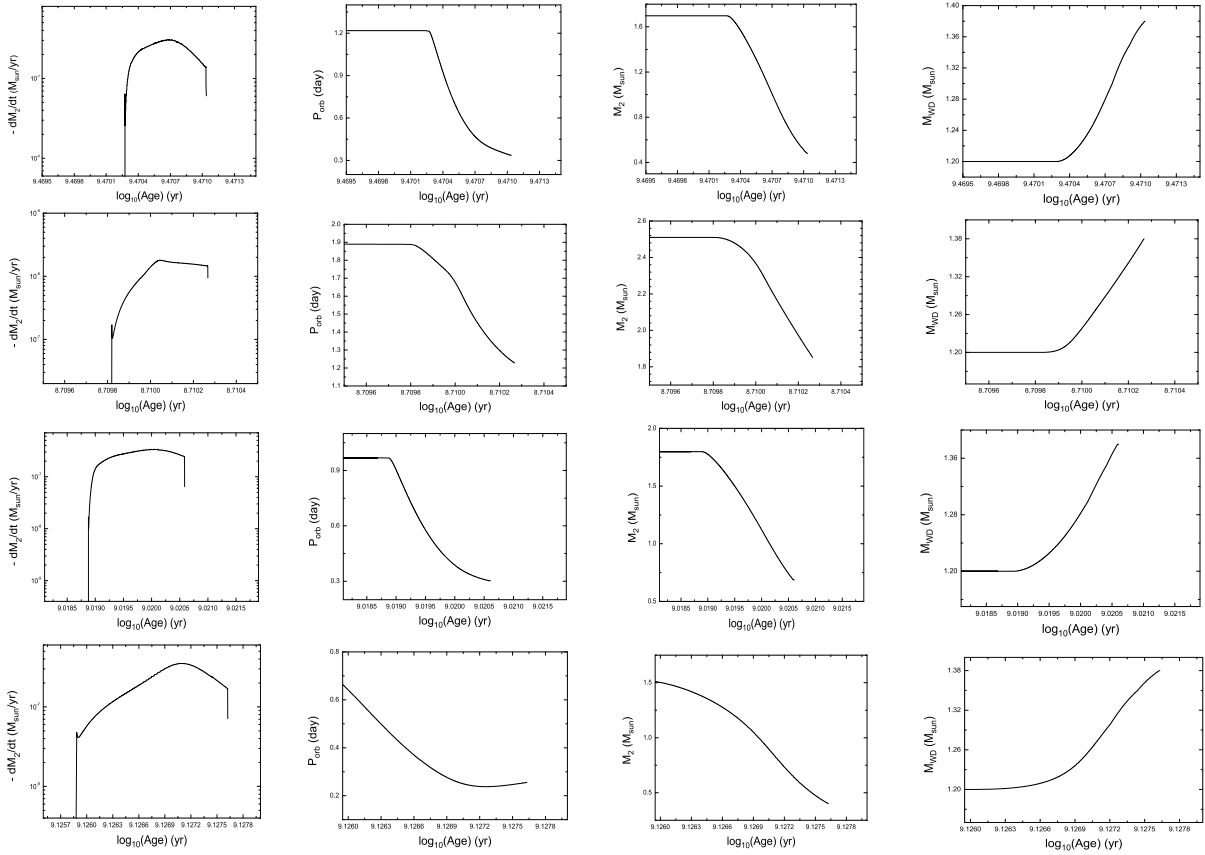}
\caption{The pre-AIC evolution of the binaries initially consisting
of a $1.2M_\sun$ WD and an MS companion star. From top to bottom
other parameters are as follows: $M_{\rm 2,i}=1.7M_\sun$ and $P_{\rm
orb,i}=1.25$ days; $M_{\rm 2,i}=2.5M_\sun$ and  $P_{\rm orb,i}=1.95$
days; $M_{\rm 2,i}=1.8M_\sun$ and  $P_{\rm orb,i}=1.0$ day; $M_{\rm
2,i}=1.5M_\sun$ and  $P_{\rm orb,i}=0.75$ days.  From left to right
are shown the mass transfer rate, the orbital period, the companion
mass, and the WD mass versus the age. } \label{fig:subfig}

\end{figure}

\clearpage

\begin{figure}
\centering
\includegraphics[scale=0.43]{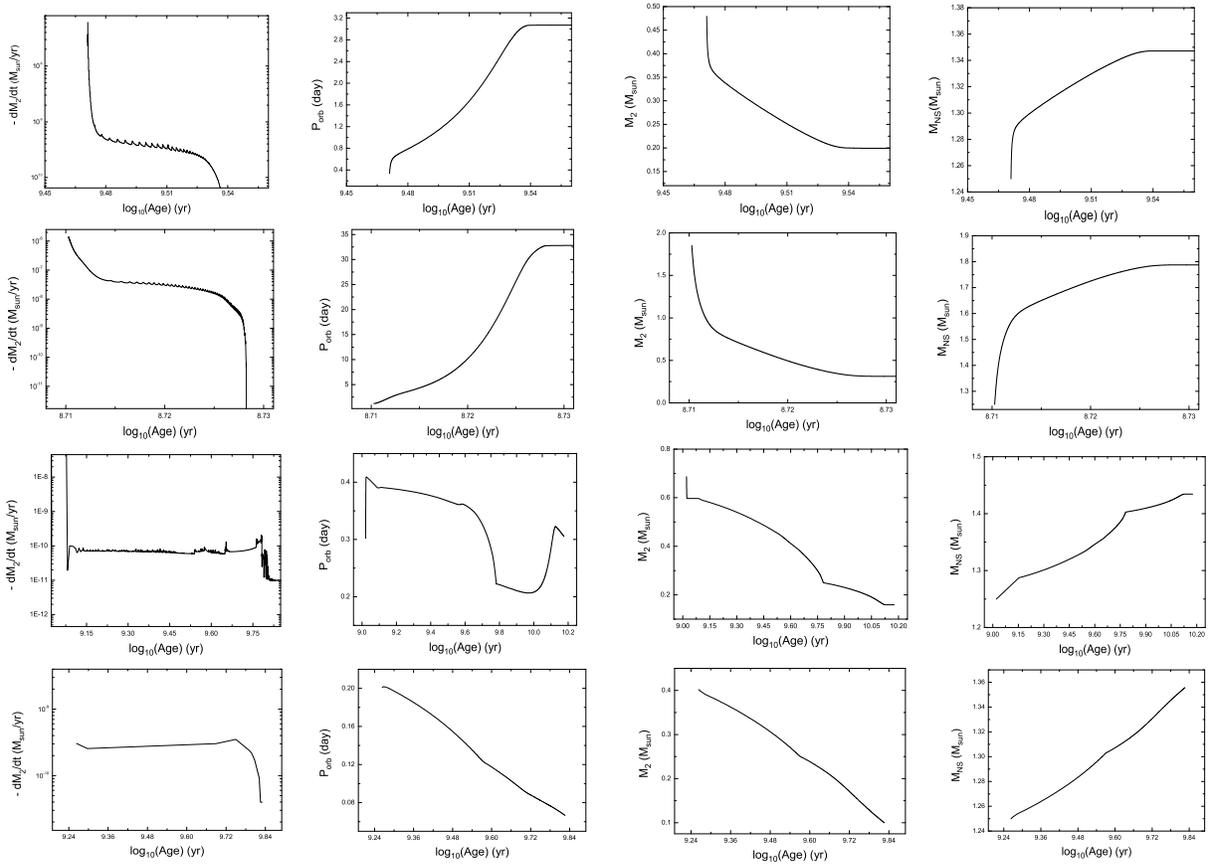}
\caption{The post-AIC evolution of the four binaries depicted in
Fig.~3.  From left to right are shown the mass transfer rate, the
orbital period, the companion mass, and the NS mass versus the age.
 }\label{fig:subfig}

\end{figure}

\clearpage

\begin{figure}
\centering
\includegraphics[totalheight=4.5in,width=5.5in]{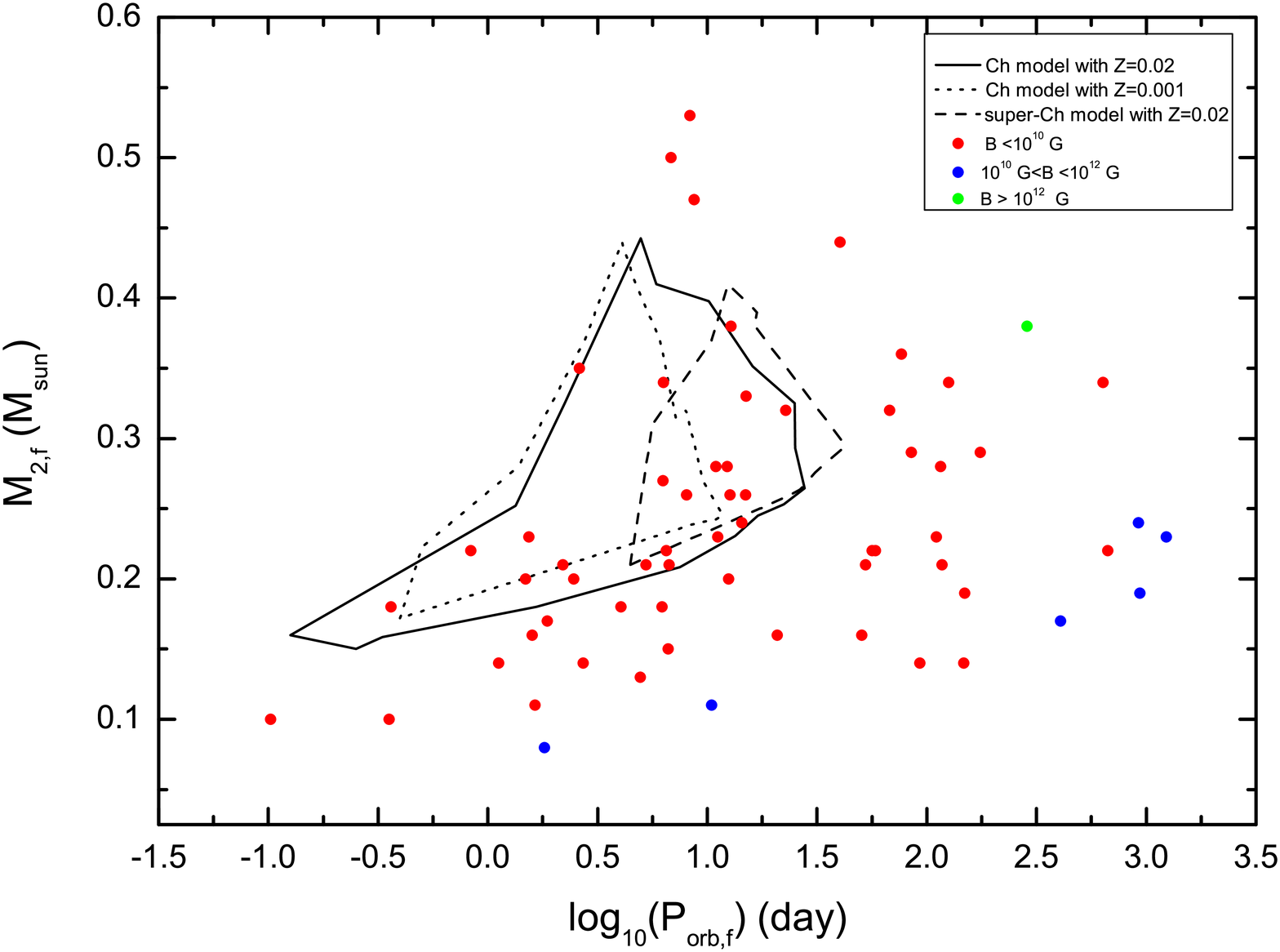}
\caption{The distributions of the final orbital periods and
companion masses of the binaries that end as PSR/WD binaries. The
solid, dotted, and dashed lines are for cases of the Chandrasekhar
model with $Z = 0.001$ and $Z = 0.02$, and the super-Chandrasekhar
model with $Z = 0.02$, respectively. The circles represent the
observed binary pulsar systems with known magnetic
fields.}\label{fig:1}
\end{figure}

\begin{figure}
\centering
\includegraphics[totalheight=5.5in,width=5.5in]{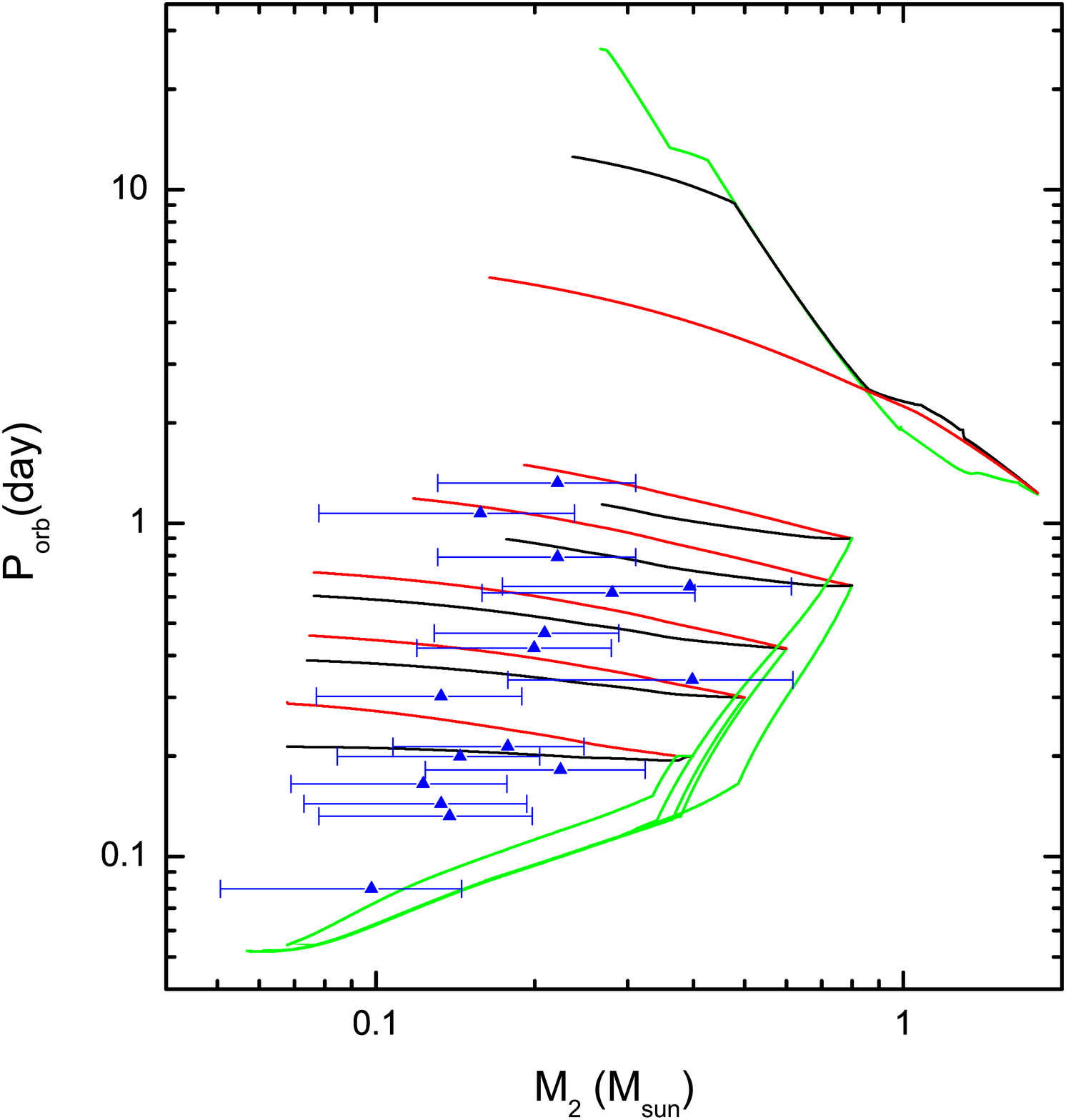}
\caption{The post-AIC evolution of six binary pulsars. The NSs are
assumed to be born as MSPs and able to ablate the secondaries. The
green, black, and red lines describe the results with the efficiency
factor $f=0.1$, 0.45, and 0.8, respectively. The known redbacks are
plotted in triangles. Their mass error bars correspond to the
orbital inclinations between $25.8\degr$ and $90\degr$ \citep[data
are taken from][and references therein]{s14a}. }\label{fig:1}
\end{figure}

\clearpage

\begin{table}

\begin{center}
\caption{Selected examples of the evolutionary sequences that form
binary pulsars. The three parts correspond to the results of the
Chandrasekhar model with $Z=0.02$ and $0.001$, and the
super-Chandrasekhar  model with $Z=0.02$, respectively.}

  \begin{tabular}{l l l l l l l l}
 \hline\hline
$\rm P_{\rm orb,i} (days)$  & $\rm M_{\rm 2,i} (\rm M_\sun)$ & $\rm P_{\rm orb,aic} (days,)$ & $\rm M_{\rm 2,aic} (\rm M_\sun)$ & $\rm P_{\rm orb,f} (days)$ & $\rm M_{\rm 2,f} (\rm M_\sun)$ & $\Delta M (\rm M_\sun)$\\
\hline
 1.2 & 1.7 & 0.32 & 0.37 & 1.56 & 0.187 & 0.06\\
 1.2 & 1.9 & 0.42 & 0.86 & 8.15 & 0.225 & 0.22\\
 1.2 & 2.4 & 0.84 & 1.85 & 28.3 & 0.275 & 0.55\\
1.95 & 2.5 & 1.23 & 1.847 & 32.81 & 0.312 & 0.537\\
1.95 & 3.2 & 0.75 & 1.1 & 6.91 & 0.377 & 0.253\\
2.9 & 3.2 & 1.22 & 0.99 & 7.63 & 0.411 & 0.202\\
2.9 & 3.5 & 1.4 & 0.738 & 4.58 & 0.428 & 0.108\\
3.5 & 3.5 & 1.93 & 0.679 & 4.97 & 0.44 & 0.083\\
\hline
1.0 & 2.2 & 0.56 & 1.5 & 17.96 & 0.25 & 0.43\\
1.0 & 2.7 & 0.61 & 2.02 & 24.62 & 0.26 & 0.61\\
2.0 & 1.9 & 0.81 & 0.975 & 17.42 & 0.25 & 0.253\\
2.0 & 2.2 & 1.31 & 1.615 & 38.54 & 0.286 & 0.465\\
2.0 & 2.9 & 0.84 & 1.625 & 15.83 & 0.347 & 0.447\\
3.0 & 2.2 & 1.65 & 1.508 & 42.6 & 0.295 & 0.424\\
3.0 & 3.0 & 1.2 & 1.42 & 15.32 & 0.38 & 0.363\\
4.0 & 3.1 & 1.73 & 1.065 & 12.41 & 0.408 & 0.229\\
\hline
1.0 & 2.0 & 0.36 & 0.68 & 5.84 & 0.213 & 0.16\\
1.0 & 2.6 & 0.46 & 1.049 & 11.73 & 0.234 & 0.28\\
2.0 & 2.1 & 1.01 & 0.627 & 9.5 & 0.26 & 0.13\\
2.0 & 2.7 & 1.54 & 0.54 & 6.08 & 0.317 & 0.08\\
2.0 & 3.2 & 1.62 & 0.531 & 3.1 & 0.373 & 0.06\\
3.0 & 2.9 & 2.9  & 0.502 & 5.42 & 0.376 & 0.045\\
3.0 & 3.4 & 1.91 & 0.57 & 3.61 & 0.429 & 0.05\\
3.5 & 3.5 & 2.32 & 0.57 & 4.08 & 0.44 & 0.046\\
\hline
\end{tabular}
\end{center}
\end{table}

\label{lastpage}

\end{document}